\documentclass[12pt,preprint]{aastex}
\begin{document}
\title{Evidence for a population of beamed radio intermediate quasars}
\author{Ting-Gui Wang\altaffilmark{1,3}, Hong-Yan Zhou\altaffilmark{1,3}, Jun-Xian Wang\altaffilmark{1,3}, You-Jun Lu\altaffilmark{2}, Yu Lu\altaffilmark{1,3}}
\altaffiltext{1}{Center for Astrophysics, University of Science and Technology of 
China, Hefei, Anhui, 230026, P.R.China}
\altaffiltext{2}{Department of Astronomy, 601 Campell Hall, University of California, Berkeley, CA 94720}
\altaffiltext{3}{Joint Institute of Galaxies and Cosmology, Shanghai Observatory and University of Science and Technology of China}

\email{twang@ustc.edu.cn}

\begin{abstract}

Whether radio intermediate quasars possess relativistic jets as radio-loud 
quasars is an important issue in the understanding of the origin of radio 
emission in 
quasars. In this letter, using the two-epoch radio data obtained during 
Faint Image of Radio Sky at Twenty centimeter sky (FIRST) and NRAO VLA Sky 
Survey (NVSS), we identified 89 radio variable sources in the Sloan Digital 
Sky Survey. Among them, more than half are radio intermediate quasars 
($RL=f_{20cm}/f_{2500\AA}<250$). For all objects with available multiple 
band radio observations, the radio spectra are either flat or inverted.  The 
brightness temperature inferred from the variability is larger than the 
synchrotron-self Compton limit for a stationary source in 87 objects, 
indicating of relativistic beaming. Considering the sample selection and 
the viewing angle effect, we conclude that relativistic jets probably 
exist in a substantial fraction of radio intermediate quasars. 
\end{abstract}


\section{Introduction}

The radio relative to their bolometric luminosity in quasars is one of 
the greatest variance in the quasar's Spectral Energy Distribution 
(SED, Elvis et al. 1994). The distribution of the radio loudness, the 
ratio of radio flux to the optical one, appears bimodal with about 90\% 
being radio quiet and 10\% radio loud (Kellermann et al. 1989; 
Falcke et al. 1996a; Ivezi{\' c} et al. 2002, 2004; c.f., Hooper et al. 1995; 
White et al. 2000; Cirasuodo et al. 2003). The confirmation and the 
origin of this radio dichotomy is one of the major issues in AGN study. 
Several scenarios were proposed to explain the large range of radio 
strengths. The basic assumptions fall into two main categories: (1) 
The central engines in the two types are assumed fundamentally different. 
For example, the formation of radio jets is associated with the spin of 
black hole (Blandford 1990 and reference therein) or with very low mass 
accretion rate onto the supermassive black hole (SMBH; Boroson 2002). 
(2) The weak of radio emission in the radio quiet is attributed to the 
rapid deceleration of relativistic jets on parsec scales, such as caused 
by the strong interaction with either gas in the inner nucleus (e.g., 
Falcke et al. 1996b) or the strong radiation produced by the quasars 
(Ghissellini et al. 2004). 

The close related question is whether radio jets in radio quiet quasars 
are relativistic as in radio loud objects. However, the measurement of 
proper motion is very difficult in general due to their weak radio 
emission and in particular due to the small size of the radio source. 
Even at the VLBI resolution, the radio sources in most low redshift 
radio quiet and radio intermediate quasars remain un-resolved or at best 
are marginally 
resolved (Falcke et al. 1996b; Blundell \& Beasley 1998; Ulvestad et al. 2005). 
The brightness temperatures or their lower limits are in the 
range of 10$^{8-11}$ K, consistent with being the nucleus origin. 
Superluminal motion at sub-parsec scale has been detected in two radio intermediate quasars
 III Zw 2 (Brunthaler et al. 2002, 2005) and PG 1407+263 (Blundell, Beasley \& Bicknell 2003), suggesting that 
the radio jets in those sources be relativistic. Radio intermediate 
quasars ($RL<250$ for flat spectrum radio sources, Falcke et al. 1996a) were 
proposed to be the boosted radio quiet quasars based on the 
statistical properties of the [OIII] luminosity, their flat radio spectra, 
the high brightness temperature and the large amplitude variability 
(Miller et al. 1993; Falcke et al. 1996b; c.f., Barvainis et al. 2005). 

As for the radio loud quasars, variability is a useful method to address 
whether a relativistic jet is present. The brightness temperature derived 
from variability is usually larger than that derived from the VLBI imaging 
in radio loud quasars 
(Valtaoja et al. 2003). In this paper, we present a detailed study of radio 
variability of the SDSS quasars by using the two epoch radio data  
derived from the Faint Image of Radio Sky at Twenty centimeters (FIRST, 
Becker et al. 1995) and NRAO VLA Sky Survey (NVSS, Condon et al. 1997). 
Throughout this paper, we assume  $H_0=70$~km~s$^{-1}$~Mpc$^{-1}$, 
$\Omega_\Lambda=0.7$ and $\Omega_m=0.3$. 

\section{The Sample and Radio Variability}

The sample was selected by cross-correlating the third Data Release of 
SDSS Quasar Catalog (Schneider et al. 2005) with the FIRST catalog (version 
03Apr11, Becker et al. 1995) and NVSS catalog (Condon et al. 1998). Since we 
are interested in radio variable sources, which should contain a compact 
core on arcsec scales, we adopt a cutoff of two arcsecs in the position 
offset as the true match. As discussed in detail by several authors, this 
selection has chance coincidence of 0.1\%, and is likely complete to 95\% 
for point sources (e.g., Gregg et al. 1996). It will miss about 7\% quasars 
with complex radio morphology or lobe dominated objects (Ivezi{\' c} et al. 2002; 
Lu et al., in preparation). For NVSS sources, we adopt a cutoff of 15 arcsecs 
in the position offset between the SDSS quasar and the radio source. This 
will lead a completeness of 90\%, while the false rate remains to be 
very low (4\%). Since the NVSS covers the entire sky north of -40 declination, 
thus encompasses all sky covered by FIRST Survey. To ensure a high 
significance of detection in NVSS, we use a threshold of 5 mJy for the 
FIRST survey. Therefore, 2010 quasars with FIRST flux larger than 5 mJy 
were searched for their NVSS countparts. To complement with possible 
complex morphology, e.g., the bright lobes, the offset was increased up 
to 45", the beam size of the NVSS survey. All of the 2010 quasars have 
at least one NVSS countpart with 45". 

Due to the different beam sizes used by NVSS and FIRST, one must be cautious 
while comparing their fluxes. We adopt a conservative approach in which 
only sources with the FIRST peak flux larger than the 
integrated flux of NVSS are considered as possible variable sources. Since 
the integrated NVSS flux may include diffuse emission or weak nearby radio 
sources, sources with NVSS flux larger than FIRST may well be due to such 
contaminations. In the next step, significance of the variation between 
the two epochs for each quasar is estimated as follows:
\begin{equation}
\sigma_{var}=\frac{S^{peak}_{FIRST}-S^{int}_{NVSS}}{\sqrt{\sigma_{FIRST}^2+\sigma_{NVSS}^2+(0.05*S^{peak}_{FIRST})^2}} 
\end{equation}
where $S^{peak}_{FIRST}$ is the peak flux during the FIRST survey, and 
$S^{int}_{NVSS}$ the integrated NVSS flux, $\sigma_{FIRST}$ and 
$\sigma_{NVSS}$ the uncertainties in the correspondent FIRST and NVSS 
fluxes. The FIRST fluxes are subject to additional systematic uncertainties 
at 5\% level, which is not included in the $\sigma_{FIRST}$ (Becker et al. 
1995). Note the 3\% systematic uncertainty of NVSS flux has already 
been included in $\sigma_{NVSS}$. We use $\sigma_{var}>3$ as a threshold 
for the source variability. Images of NVSS and FIRST are then visually 
examined for possible contamination due to nearby bright sources, and 
three of them were removed for this reason from the sample. SDSS 
J094420.44+613550.1 was eliminated from the sample because of NVSS 
catalog gives a wrong flux for this object. This gives a sample of 
89 variable radio quasars with $f_{FIRST}/f_{NVSS}$ in the range of 
1.2 to 2.6. We expect that three out of them are spurious with this 
3$\sigma$ limits.

Radio loudness is calculated using the FIRST radio flux and the PSF magnitudes 
derived from the SDSS survey as follows: 
\begin{equation}
RL= (f_{1.4GHz}/f_{\nu, 2500\AA})
\end{equation}
$k$-corrections to the optical flux is derived using the five SDSS magnitudes, 
corrected for the Galactic reddening. While we assume $\alpha_r=0.0$ for 
the $k$ correction of the radio flux of radio variable sources since they 
are likely flat spectral radio sources (see below). Apparently, 67\% of 
radio variable sources are radio intermediate or radio quiet (Fig 1) 
according to the definition of Falcke et al. (1996a).  

For the sources with significant flux variation, we compute the 
the lower limit of the brightness temperature by assuming the variable 
part of radio flux is emitted in a region of which the light-crossing 
time is equal to the separation of the two observations 
\begin{equation}\label{eq5}
T_{B}^{l}\sim \frac{\Delta P_{\nu}}{2k_{B}\nu^2\Delta t^2},
\end{equation}
where $\Delta P_{\nu}$ is the variable part of the radio power
computed from the difference between the FIRST and NVSS fluxes,
$\Delta t$ the timescale of the radio flux variability estimated 
from the difference of the observation time of FIRST and NVSS in the 
source rest frame, and $k_B$ is the Boltzmann constant. 
Since $\nu\Delta t$ is Lorentz invariant, we simply use the 
observe frequency (1.4 GHz) and the separation of NVSS and FIRST 
observation dates. In cases where the FIRST observation date is only 
accurate to month, we use the day that makes the separation maximum 
to give a conservative estimate. The brightness temperatures inferred 
in this way are all larger than 10$^{12}$ K except for two low redshift 
quasars (see also Table 1).

The upper limit of the brightness temperature from the synchrotron radiation 
of a stationary source due to inverse Compton process is approximately 
$10^{12}$ K (Kellermann \& Pauliny-Toth 1969). This temperature may be 
greatly exceeded if the emission region moves relativistically towards 
the observer or if a coherent radiation process is responsible for the radio 
emission. In the former case, the apparent brightness temperature is boosted 
by a factor of
\begin{equation}
f=T^{var}_B/T^{intr}_B=D^3
\end{equation}
where $D=[\gamma*(1-\beta \cos\theta)]^{-1}$ is the Doppler factor, 
$\gamma=1/\sqrt{(1-\beta^2)}$,  $\beta=v/c$, and $\theta$ is the angle between 
the line of sight and the velocity of the jet. However, it was shown that 
real radio sources may emit at equi-partition brightness temperature around 
10$^{11}$ K in most circumstance (e.g., Readhead 1994). In this paper, we 
adopt the inverse Compton limit as a secure upper limit to estimate the 
lower limit of the Doppler factor for the sources with brightness 
temperatures greater than this limit.

The lower limits of the Doppler factors we derived are in the range 
of 0.6-25 with a median around 4. For these radio variable sources, the 
maximum brightness attained ($\sim$ 10$^{16}$ K) are similar for radio 
loud, radio intermediate quasars (\cite{TBvsRL}). Note that the lack
of sources with high RL and low brightness temperature T$_B$ in the figure
is due to a selection effect that sources with higher RL tend to be
brighter in radio, thus have much higher radio power variation $\Delta$P
based on similar fractional variation amplitudes. 

The fraction of variable sources varies with the radio power and radio 
loudness (\cite{radiopower}). The probability for a constant 
fraction at different radio power is only 2$\times 10^{-5}$ ($D=0.255$) 
using two sided 
Kolmogorov-Smornov test for two unbined distributions if an average 
radio spectral index $\alpha=0.5$ for the parent sample of radio loud 
quasars. It increases to 2\% if $\alpha=0.0$ is used. Sources appears 
most-variable for radio powers in the range of $10^{24}-10^{26.5}$~W~
Hz$^{-1}$  and least variable in $10^{26.5}-10^{28}$~W~Hz$^{-1}$. 
Quasars are more likely variable at radio loudness $<10^{2.5}$ than 
above this value. The radio loudness distributions for variable 
and parent samples are drawn from the same population at a probability of 
only $10^{-7}$ ( 1\%) using two-sided Kolmogorov-Smornov test 
for two unbinned distributions if $\alpha_r$=0.0 (0.5) is used for 
$k$-correction in radio flux. Excluding radio sources with deconvolved 
major axis of the FIRST image larger than 3 arcsecs has little effect on 
these results.

The detection rate does not change with the separation of the two epoches, 
which is in the range of 0.4 to 5 years in the source rest frame (\cite{fractdt}). This is a possible indication that the variability is dominated 
by flares, similar to what observed in III Zw 2, but rather than long term 
smooth variations. 


With the Doppler factor, we estimate the maximum viewing angle between 
the line of sight and the jet assuming an intrinsic narrow jet as follows, 
\begin{equation}
\cos\theta_0=\min ([1-\sqrt{(1-\beta^2)}/D]/\beta) \;\;\; {\mathrm for\; all \; \beta \le 1 }
\end{equation}
The inferred maximum viewing angles are quite small (usually, 
$\le$15$^{\mathrm o}$) for most radio variable sources. And in some 
extreme objects, this angle is even smaller than the opening angle of 
jets on parsec scales or kpc scales in nearby radio galaxies  
(typical of 5$^{\mathrm o}$, e.g., Ly, Walker \& Wrobel 2004). 

Four quasars showed variations of a factor more than two between 
NVSS and FIRST surveys. Three of them are radio intermediate even 
at their peak radio flux during the FIRST survey.  Their radio powers at 
20cm are moderate ($24.9<\log~P_{\rm 20cm} <25.9$). The 
fastest variation among them is a factor of 2.5 in 20 
cm flux within eight months for SDSS J073938.85+305951.2. 
We examined the optical
spectra for possible evidence of blazar-like feature but fails to find 
any. In particular, the line and continuum spectra appear similar 
to the composite quasar spectra (VanDen Berk et al. 2001).

Some interesting objects are noted. This sample contains several radio 
loud BAL QSOs, which were analyzed in detail in Zhou et al. (2005).
SDSS 094857.31+002225.5 is an extremely radio loud NLS1 with prominent optical 
FeII emission and inverted radio spectrum. It was studied in detail by 
Zhou et al. (2003). They proposed a relativisitic jet in this object 
based on the flux variability between the FIRST and NVSS data. It was 
observed by VLBA in 2 and 8.4GHz, and remains unresolved (Beasley et al. 2002), 
consistent with the conclusion reached by Zhou et al.. With an inferred 
Doppler factor of $>$4.3, the intrinsic radio loudness of this object is in 
the radio intermediate range if most of the radio flux is contributed by 
the beaming component. 

For 10 radio variable objects that the non-simultaneous multi-wavelength 
radio observations are available, the radio spectral indices are in the 
range of $-0.2\le\alpha\le 0.6$ ($f_\nu\propto \nu^\alpha$) with a median 
of 0.2. This is also consistent with boosted radio emission.
 
\section{Discussion}

Among 2010 quasars with FIRST flux larger than 5 mJy, we identified 89 radio 
variable ones. For strong sources, the systematic uncertainty limits any 
variation detectable to $\gtrsim$20\% at 3$\sigma$ levels (5\% systematic 
uncertainty in the 
FIRST flux and 3\% in the NVSS flux). At lower radio flux, the statistical 
fluctuation is significant, the limit increases to $\simeq 30$\% at flux limit 
of 5 mJy. Considering that we counted only sources with 
$f_{FIRST}>f_{NVSS}$, about 9\% quasars with radio flux larger than 5 mJy 
show variations in 20 cm radio flux at 20-30\% level on time scale of years.

We have found strong evidence for relativistic beaming in 87 of these 89 
radio variable quasars. Among them, two are radio quiet ($RL<10$) and 29 
are radio loud ($RL>250$), while the majority (56) are radio intermediate. 
Since radio powers of most (70\%) these radio intermediate quasars are above 
the break power between FR I and FR II division ($\log P_{20cm}{\mathrm 
(W~Hz^{-1})} \simeq 25.0$), one may question whether they are true radio 
intermediate quasars.  
However, radio powers of most these quasars are boosted by relativistic effect 
and as such represent only upper limits. The intrinsic radio power may be 
much lower. For example, if we use the lower limits of Doppler factor 
derived from the radio variability to de-boost the radio flux, the
radio powers for these sources would be a factor of 10-1000 lower
\footnote{Radio flux from jet has been boosted by a factor of 
$D^{2+\alpha}$ for an optically thin jet model and  $D^{3+\alpha}$ for 
isotropic emission model.}, well in the range for FR I galaxies. Of 
course this correction is 
oversimplified considering the contribution from lower velocity part of 
the jet and extended lobes, and the true powers may lie in between. By 
noting that the radio powers of most objects are not too far from the 
break power, even if substantial correction by the boosting effect is 
introduced, most of these quasars would be in the FR I regime. Therefore, 
we believe that most of objects with $RL<250$ are radio intermediate as 
proposed by Fackle et al.. This is hitherto the largest sample of radio 
intermediate quasars with beamed radio emission. 



Among 2010 radio quasars with 20 cm flux above 5 mJy, 724 objects show radio 
loudness $RL<250$. Since radio spectral indexes are not available for all these 
sources, we can only set an upper limit on the number of radio intermediate 
quasars in the sample to 724. Among them we detected 58 quasars with 
relativistic jets beaming toward us. This leads to an apparent fraction to 
$\gtrsim$8.0\%. However, a number of serious corrections must be applied in 
order to estimate the fraction of quasars with relativistic jets. 

First, because of large inferred Doppler factor for these 58 radio 
intermediate/quiet quasars, they must be observed very close to the 
direction of jet. We estimate that the line of sight is within 
15$^{\mathrm o}$ of the jet direction for most variable objects 
({\rm Eq.} 2), thus the probability of detecting such a source is small. 
This implies that their parent population may be large. Second, since 
the quasars in the variable sample are relativistically boosted, the flux 
limit of the parent population may be well below 5 mJy, which is used 
to select the sample.  Both effects depends on the Doppler factor: 
sources with a larger Doppler factor are seen at a small solid angle around 
the jet direction, and their fluxes are boosted by a larger factors. 
Using the maximum extending angles estimated in the last section, we estimate 
that the minimum size of parent quasar population at an opening angle less 
than $\theta$ (or with a given $D$) using:
\begin{equation} 
N(>D)=\frac{{\rm number\; of\; objects\; with } \cos\theta_{0}\leq\cos\theta}{(1-\cos \theta)}, 
\end{equation}
where $\cos\theta$ is the $\cos\theta_0$ for Doppler factor $D$ ($Eq.$ 5). 
The result is plotted in \cite{parentpop}. The number of parent sources 
increases with the Doppler factor. This is in fact fully consistent with 
the assumption that the radio flux in these variable sources are boosted 
greatly by relativistic effect, as such objects with larger Doppler factor 
trace a larger parent population with a lower flux limit. At Doppler factor 
larger than 5, the radio variable sources traces a parent population 
$\approx$ 1500-2000. The number should be doubled since we consider only these 
sources with FIRST flux larger than the NVSS one. This number is a factor 
of 4 larger than all radio intermediate quasars at flux limit of 5 mJy 
even we relax the radio loudness of radio intermediate quasars to $RL<250$ 
for all quasars. 

The ratio of jet component at its comoving frame to the isotropic 
component is needed to determine the flux limit of parent population 
with a given Doppler factor, which is required to estimate the number 
of parent population. However, this is not constrained even for the 
best studied radio intermediate quasar, III Zw 2. In that case, the 
observed flux for extended lobe is 10-20\% of the core component, and 
apparent velocity is 2.6c. Since we do not know the angle between the 
line sight to the jet, it Doppler factor cannot be determined. 
Fortunately, a meanful conclusion can still be reached with the current 
data. If the radio emission of quasars with $D\simeq 5$ have been boosted 
by a factor of 10, then the flux limit of parent population is 0.5 mJy for 
our sample. 
%
%
%
Note that to the FIRST flux limit (~1 mJy), 4124 (5313) among 44984 
quasars in SDSS DR3 are detected in the FIRST with 3 arcsec offset in 
the flux limit of $\sim$ 1 mJy ($\sim$0.7 mJy) (de Vries, Becker \& White 
2005). Most of them are radio intermediate (see Iveciz et al. 2003). 
Extrapolating to 0.5mJy will predict the number of radio detectable quasars 
to 6751. With this assumption, the parent population of the radio variable 
sources would be as large as half of quasars with radio flux larger than 
0.5 mJy. Alternatively, if the isotropic component (lobes) is very weak 
for these radio intermediate quasars, then the flux limit of the parent 
population is even lower, we will detect a small fraction of relativistic 
jets but intrinsically radio even weak quasars.

Note this number is likely a conservative lower limit because not all 
quasars with relativistic jets shows radio variations at amplitude greater 
than our detection limits during the epoch of two radio observations.
Thus we believe that relativistic jets present in most radio intermediate 
quasars.
  
Our results imply that jets in a large fraction of radio intermediate 
sources are relativistic, but the size of emission region is smaller than 
in classical radio loud quasars. A substantial fraction of 20 cm radio flux 
 is emitted on the scales of several to tens parsecs as suggested by the 
variability time scale. The peak brightness or the Lorentz factor of the 
most compact component is similar for both radio loud and radio intermediate 
quasars. The typical size of radio loud objects, however, appears larger 
given their large radio powers. As a result, the 20 cm radio emission in 
the classical radio sources is less variable on time scales of years. This 
is consistent with our results that the fraction of radio variable source 
is small at very large radio loudness.

The presence of relativistic jets on parsec scales in a substantial fraction 
of radio intermediate quasars has several implications for AGN models. If 
the radio emission has been boosted by relativistic effect and the emission 
from extended lobes are weak as of III Zw 2, many flat spectral intermediate 
radio quasars might be boosted radio-quiet quasars. If the broad band spectrum 
of jet component is similar to that of low peak BL Lacs while the disk 
emission component is similar to that of other quasars, we estimate that 
the jet emission will dominate the SED at sub-mm wavelength, and contribute 
significantly to the hard X-ray spectrum of those beamed objects. 
 
\acknowledgments We wish to thank the referee for constructive comments. 
This work was supported by Chinese NSF through NSF 10233030 and NSF 
10573015, the Bairen Project of CAS.

\clearpage

\begin{figure}
\epsscale{1.0}
\plotone{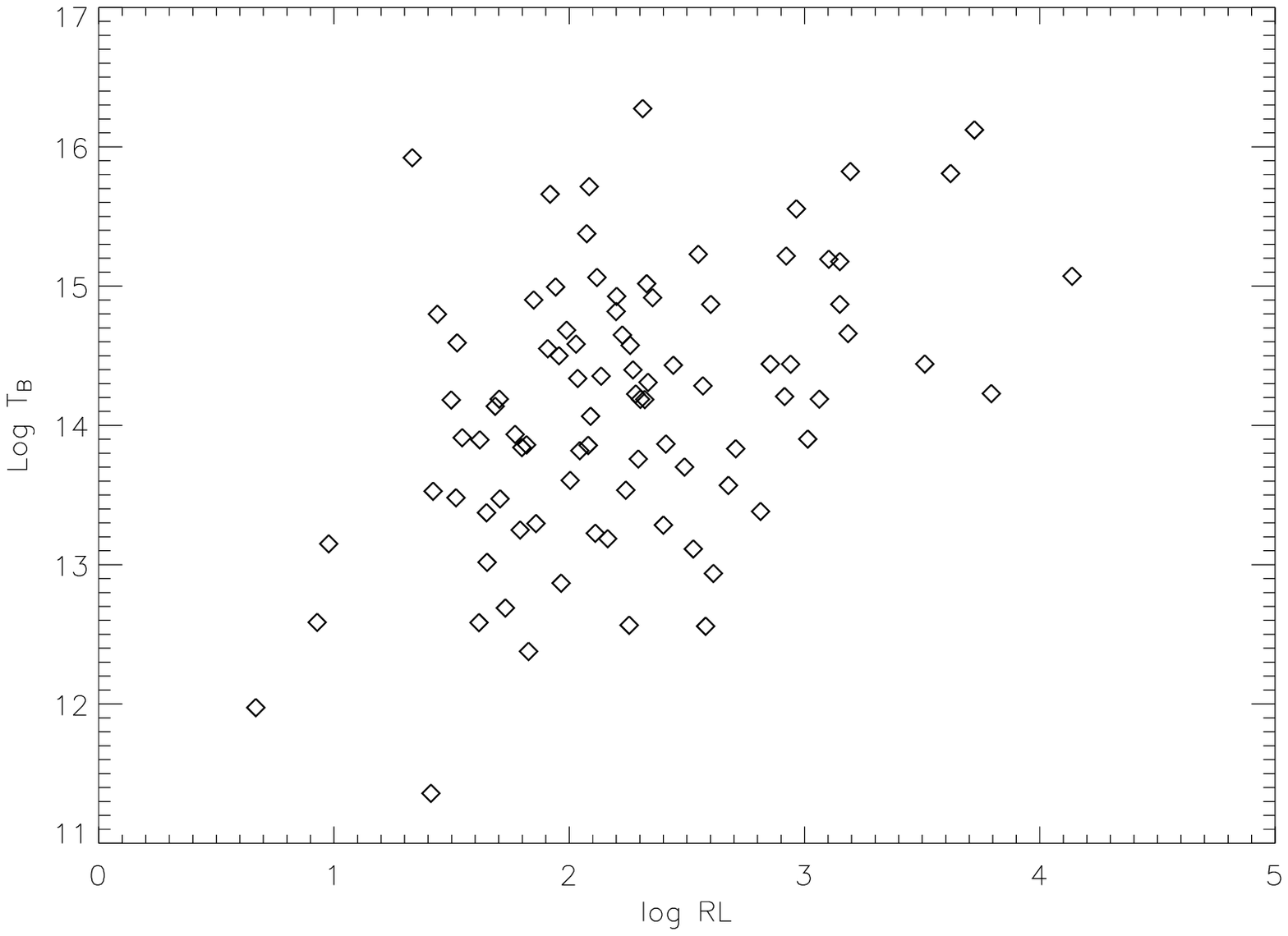}
\caption{The brightness temperature versus radio loudness for 89 radio variable quasars.}\label{TBvsRL}
\end{figure}

\clearpage

\begin{figure}
\epsscale{1.0}
\plottwo{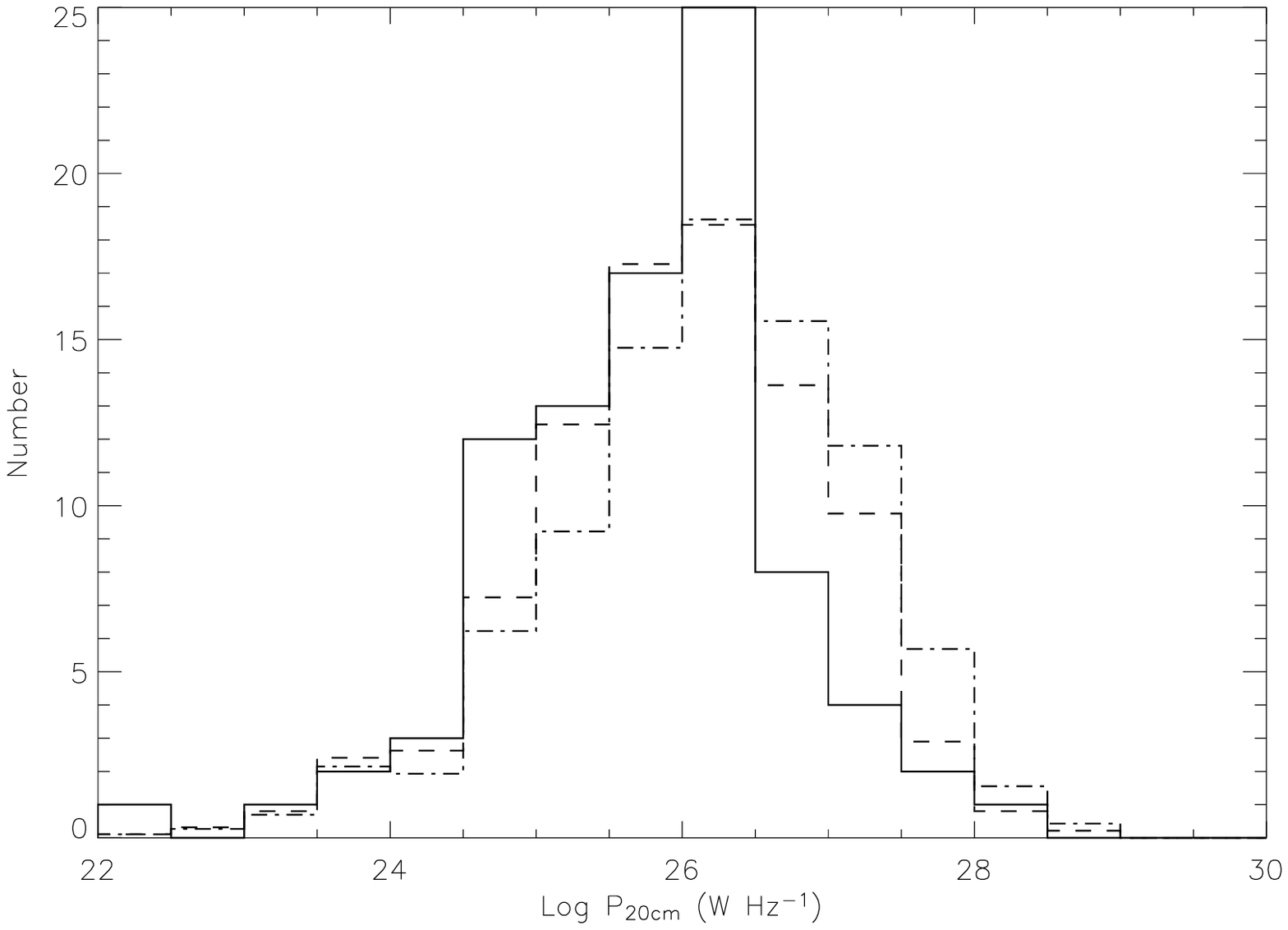}{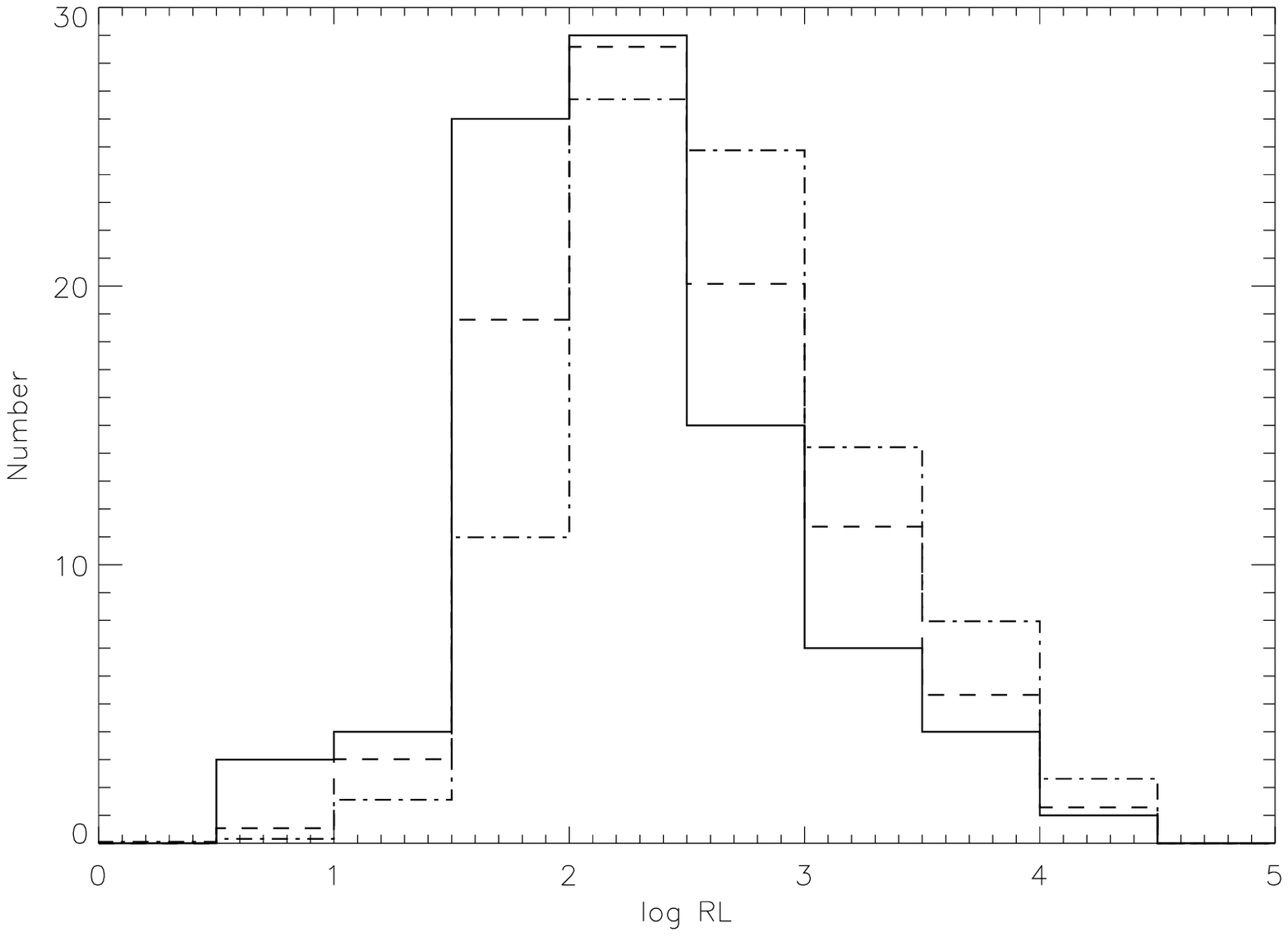}
\caption{The distribution of radio loudness (left panel) and radio power (right panel) for radio variable quasar 
sample (solid line) and the parent radio selected quasar sample (dashed 
line: k-correction with $\alpha_r=0$, dash-dot line: $\alpha_r=0.5$). 
The number of quasars in the parent sample is normalized to the former. 
}\label{radiopower}
\end{figure}

\clearpage

\begin{figure}
\epsscale{1.0}
\plotone{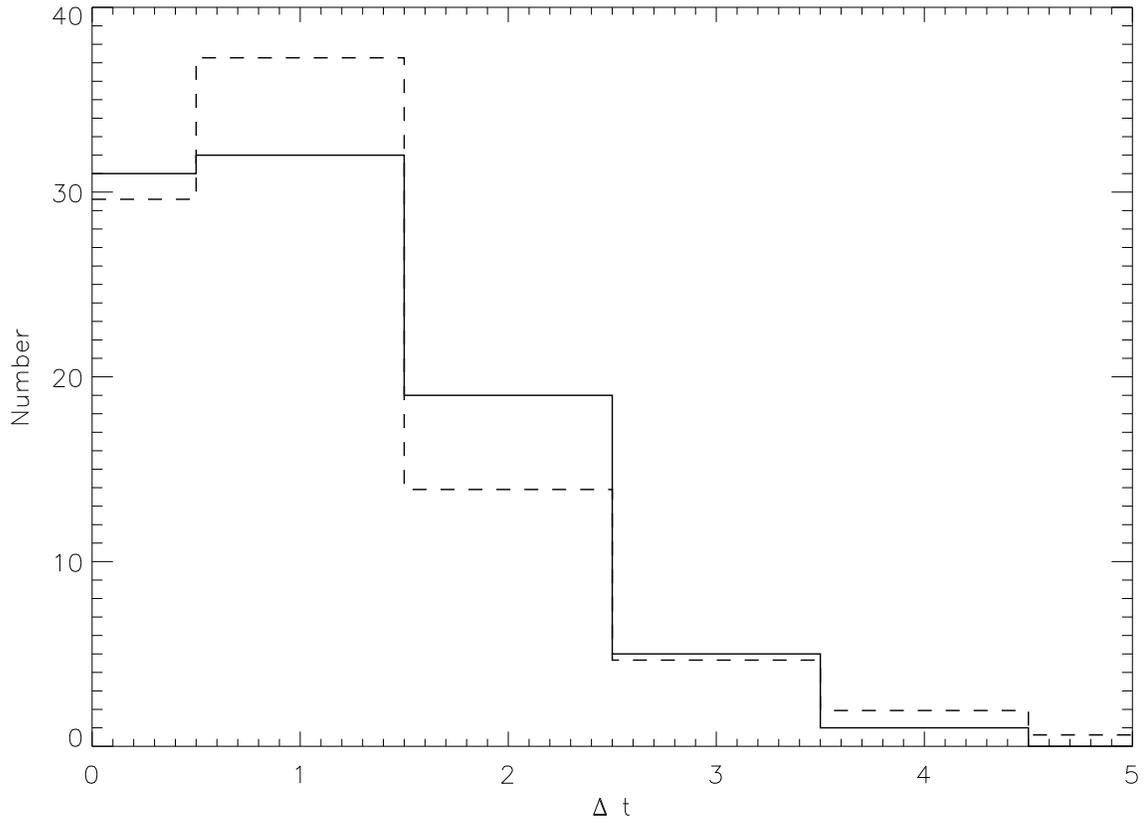}
\caption{The distribution of the separation between FIRST and NVSS 
observations for the variable sample and the parent sample.}
\label{fractiondt}
\end{figure}

\clearpage

\begin{figure}
\epsscale{1.0}
\plotone{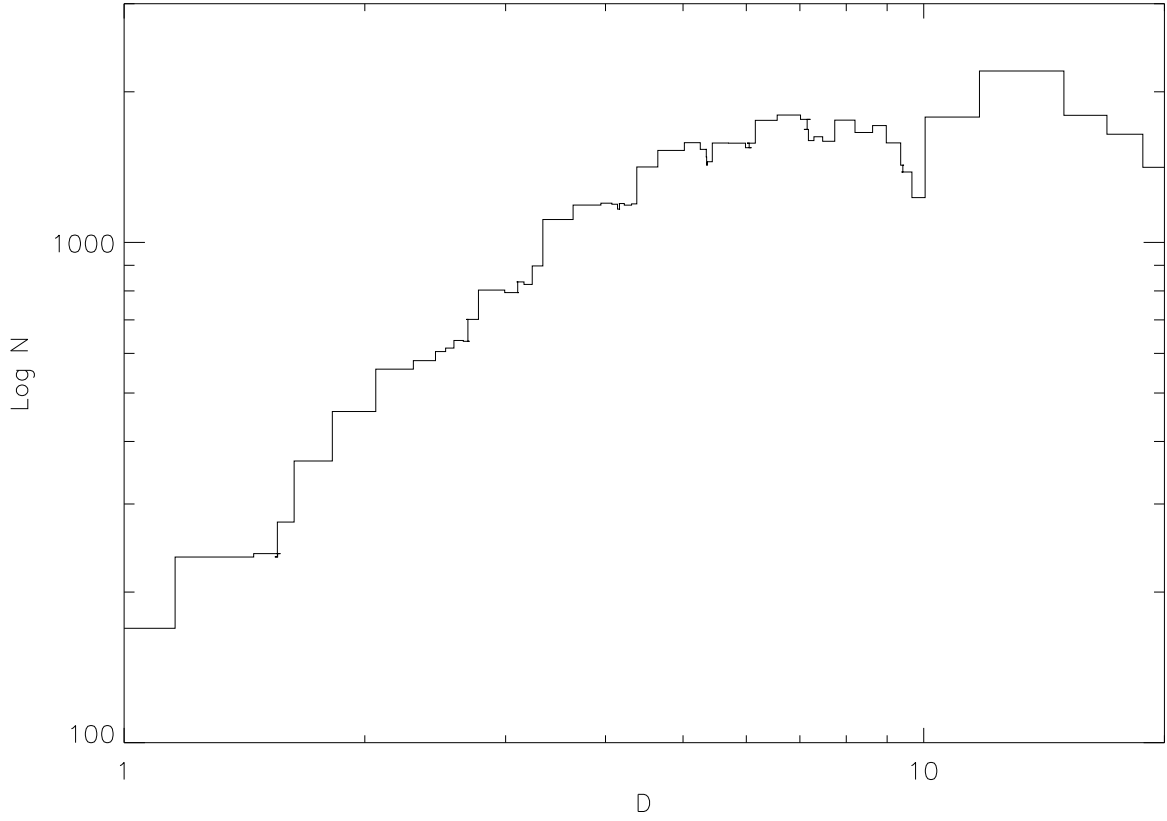}
\caption{The expected number of parent population versus Doppler factor (See {\rm Eq.} 6). 
}\label{parentpop}
\end{figure}

\clearpage

\begin{deluxetable}{rrrrlrlrrr}
\setlength{\tabcolsep}{0.02in} \tablecaption{Basic data of Radio Variable Quasars\label{t1}} 
\tabletypesize{\scriptsize}
\tablehead{ \colhead{Coordinates (J2000)} & \colhead{Redshift} & \colhead{$g'$} &
\colhead{$S_{FIRST}^{p}$} & \colhead{FIRST Date} & \colhead{$S_{NVSS}^{int}$} & \colhead{NVSS Date}
 &\colhead{$\log RL$} & \colhead{$\log T_{B}$} & \colhead{$D$} \\
\colhead{hhmmss.ss$\pm$ddmmss.s} & \colhead{} & \colhead{mag} & \colhead{mJy} & \colhead{yyyy-mm-dd} &
\colhead{mJy} & \colhead{yyyy-mm-dd} &\colhead{} & \colhead{K} &\colhead{}  
}
\startdata
010249.65$-$085344.4 & 1.6825 & 18.484$\pm$0.014 &   8.35$\pm$ 0.15 & 1997-02-   &   5.70$\pm$ 0.50 &1993-09-20 & 1.54 & 13.91 &  4.34\\
013146.43$-$084104.1 & 1.6512 & 19.488$\pm$0.023 &   5.47$\pm$ 0.15 & 1997-02-   &   3.00$\pm$ 0.50 &1993-09-20 & 1.82 & 13.86 &  4.17\\
015153.30$-$002850.2 & 1.9950 & 18.076$\pm$0.016 &  12.14$\pm$ 0.15 & 1995-11-14 &   9.20$\pm$ 0.50 &1993-11-15 & 1.52 & 14.59 &  7.31\\
073938.85+305951.2 & 3.3994 & 21.256$\pm$0.040 &   7.26$\pm$ 0.37 & 1993-04-12 &   2.80$\pm$ 0.40 &1993-12-15 & 2.31 & 16.27 & 26.58\\
074033.54+285247.1 & 0.7111 & 19.376$\pm$0.020 &  77.57$\pm$ 0.15 & 1993-04-29 &  63.30$\pm$ 1.90 &1993-12-15 & 3.15 & 15.18 & 11.44\\
074815.44+220059.5 & 1.0595 & 16.437$\pm$0.015 &   9.86$\pm$ 0.13 & 1998-08-   &   7.00$\pm$ 0.40 &1993-11-01 & 0.98 & 13.15 &  2.42\\
074823.85+332051.2 & 2.9888 & 20.036$\pm$0.021 &   8.14$\pm$ 0.14 & 1995-10-   &   5.90$\pm$ 0.40 &1993-12-15 & 1.94 & 14.99 &  9.95\\
075535.41+292047.3 & 0.5064 & 18.552$\pm$0.115 &   7.19$\pm$ 0.15 & 1993-05-04 &   5.40$\pm$ 0.40 &1993-12-15 & 1.77 & 13.94 &  4.42\\
075849.47+305452.8 & 2.7967 & 19.720$\pm$0.020 &   5.92$\pm$ 0.14 & 1993-04-13 &   4.20$\pm$ 0.40 &1993-12-15 & 1.92 & 15.66 & 16.60\\
080945.16+453918.0 & 2.0395 & 20.708$\pm$0.030 &  20.59$\pm$ 0.13 & 1997-03-18 &  15.10$\pm$ 0.60 &1993-11-15 & 2.85 & 14.44 &  6.51\\
081352.87+352035.4 & 1.8981 & 19.622$\pm$0.019 &   8.01$\pm$ 0.13 & 1994-07-03 &   4.70$\pm$ 0.40 &1993-12-15 & 2.08 & 15.71 & 17.30\\
081415.05+412323.4 & 1.2969 & 18.487$\pm$0.025 &   5.15$\pm$ 0.13 & 1994-09-02 &   3.40$\pm$ 0.40 &1993-12-15 & 1.44 & 14.80 &  8.57\\
081655.28+475611.5 & 2.2337 & 20.293$\pm$0.025 &   9.01$\pm$ 0.47 & 1997-04-05 &   4.90$\pm$ 0.40 &1993-11-15 & 2.27 & 14.40 &  6.30\\
082817.25+371853.7 & 1.3530 & 21.675$\pm$0.056 &  21.18$\pm$ 0.13 & 1994-07-23 &  14.80$\pm$ 0.60 &1993-12-15 & 2.97 & 15.55 & 15.30\\
083225.34+370736.2 & 0.0919 & 16.173$\pm$0.014 &  11.78$\pm$ 0.17 & 1994-07-23 &   8.20$\pm$ 0.50 &1993-12-15 & 0.93 & 12.59 &  1.57\\
083655.73+342335.4 & 0.7905 & 20.258$\pm$0.025 &  10.32$\pm$ 0.13 & 1994-07-01 &   6.30$\pm$ 0.40 &1993-12-15 & 2.60 & 14.87 &  9.04\\
083658.91+442602.3 & 0.2544 & 15.613$\pm$0.027 &   9.39$\pm$ 0.15 & 1997-02-28 &   6.60$\pm$ 0.50 &1993-11-15 & 0.67 & 11.97 &  0.98\\
083744.01+420643.9 & 2.1301 & 19.337$\pm$0.024 &  18.12$\pm$ 0.13 & 1995-12-19 &  13.50$\pm$ 0.60 &1993-11-15 & 2.20 & 14.82 &  8.69\\
083951.00+333630.9 & 1.7528 & 19.953$\pm$0.023 &   5.60$\pm$ 0.14 & 1994-06-19 &   4.00$\pm$ 0.40 &1993-12-15 & 2.07 & 15.38 & 13.35\\
084955.28+005305.5 & 1.0345 & 18.639$\pm$0.021 &   6.20$\pm$ 0.15 & 1998-08   &   4.00$\pm$ 0.40 &1993-11-15 & 1.65 & 13.02 &  2.18\\
084957.97+510829.0 & 0.5837 & 18.956$\pm$0.021 & 344.09$\pm$ 0.14 & 1997-04-25 & 266.30$\pm$ 8.00 &1993-11-15 & 3.79 & 14.23 &  5.53\\
085001.17+462600.5 & 0.5238 & 19.137$\pm$0.022 &  20.90$\pm$ 0.13 & 1997-03-22 &  16.00$\pm$ 0.60 &1993-11-15 & 2.61 & 12.94 &  2.05\\
085217.84+054027.8 & 0.8520 & 20.915$\pm$0.035 &   5.45$\pm$ 0.15 & 2000-02-   &   3.30$\pm$ 0.40 &1993-11-15 & 2.58 & 12.56 &  1.54\\
085958.69+455237.9 & 0.4400 & 18.860$\pm$0.020 &  30.74$\pm$ 0.14 & 1997-03-22 &  19.60$\pm$ 0.70 &1993-11-15 & 2.53 & 13.11 &  2.35\\
090111.86+044858.8 & 1.8626 & 19.526$\pm$0.023 & 133.57$\pm$ 0.15 & 2000-02-   &  94.30$\pm$ 2.90 &1993-11-15 & 3.18 & 14.66 &  7.70\\
090155.15+425404.4 & 1.7350 & 19.518$\pm$0.021 &  14.23$\pm$ 0.14 & 1997-02-17 &   9.90$\pm$ 0.50 &1993-11-15 & 2.32 & 14.19 &  5.35\\
090412.87+060326.5 & 0.9360 & 18.171$\pm$0.024 &   8.01$\pm$ 0.15 & 2000-02-   &   6.20$\pm$ 0.40 &1993-11-15 & 1.62 & 12.58 &  1.57\\
090743.66+551512.4 & 0.6448 & 17.409$\pm$0.014 &  22.58$\pm$ 0.14 & 1997-03-   &  16.80$\pm$ 0.60 &1993-11-23 & 1.79 & 13.25 &  2.61\\
091641.76+024252.8 & 1.1025 & 19.121$\pm$0.016 &  99.39$\pm$ 0.14 & 1998-07-   &  72.50$\pm$ 2.20 &1993-11-15 & 3.06 & 14.19 &  5.36\\
093215.14+432738.4 & 0.9564 & 18.237$\pm$0.052 &  20.43$\pm$ 0.13 & 1997-02-20 &  12.30$\pm$ 0.50 &1993-11-15 & 2.04 & 13.82 &  4.04\\
093323.02$-$001051.6 & 0.7949 & 18.613$\pm$0.014 & 101.36$\pm$ 0.15 & 1998-08-   &  66.80$\pm$ 2.00 &1995-02-27 & 2.92 & 14.21 &  5.44\\
093818.35+390809.8 & 1.3049 & 20.423$\pm$0.028 &   7.43$\pm$ 0.13 & 1994-08-13 &   5.50$\pm$ 0.50 &1993-12-15 & 2.35 & 14.92 &  9.38\\
094857.31+002225.5 & 0.5846 & 18.661$\pm$0.012 & 107.53$\pm$ 0.15 & 1998-09-   &  69.50$\pm$ 2.10 &1995-02-27 & 3.01 & 13.90 &  4.30\\
095046.47+584113.0 & 2.3648 & 20.426$\pm$0.025 &   6.54$\pm$ 0.14 & 2002-06-   &   4.80$\pm$ 0.40 &1993-11-23 & 2.40 & 13.28 &  2.68\\
095147.86+020235.5 & 0.6053 & 18.424$\pm$0.024 &   6.14$\pm$ 0.15 & 1998-07-   &   4.20$\pm$ 0.50 &1995-02-27 & 1.73 & 12.69 &  1.70\\
095227.30+504850.6 & 1.0909 & 17.848$\pm$0.032 & 104.85$\pm$ 0.15 & 1997-04-25 &  85.80$\pm$ 3.00 &1993-11-15 & 2.57 & 14.28 &  5.77\\
095618.17+542628.2 & 1.7147 & 19.169$\pm$0.021 &   8.72$\pm$ 0.14 & 1997-05-   &   6.40$\pm$ 0.50 &1993-11-23 & 2.08 & 13.86 &  4.16\\
095739.92+074047.9 & 1.6688 & 19.454$\pm$0.023 &  73.56$\pm$ 0.15 & 2000-01-   &  54.70$\pm$ 1.70 &1995-02-27 & 2.94 & 14.44 &  6.50\\
095819.66+472507.8 & 1.8818 & 18.545$\pm$0.025 & 763.01$\pm$ 0.15 & 1997-03-31 & 603.80$\pm$18.10 &1993-11-15 & 3.62 & 15.81 & 18.59\\
101609.48+002810.5 & 1.0131 & 19.891$\pm$0.022 &  22.47$\pm$ 0.15 & 1998-08-   &  18.10$\pm$ 0.70 &1995-02-27 & 2.68 & 13.57 &  3.34\\
103424.41+493221.0 & 1.6163 & 20.105$\pm$0.020 &  12.42$\pm$ 0.13 & 1997-04-17 &   9.70$\pm$ 0.50 &1993-11-15 & 2.41 & 13.87 &  4.19\\
104901.71+005534.0 & 1.1633 & 18.408$\pm$0.031 &   5.73$\pm$ 0.15 & 1998-08-   &   3.20$\pm$ 0.40 &1995-02-27 & 1.52 & 13.48 &  3.12\\
105320.42$-$001649.6 & 4.3032 & 21.952$\pm$0.085 &  13.31$\pm$ 0.15 & 1998-08-   &   9.30$\pm$ 0.50 &1995-02-27 & 2.12 & 15.06 & 10.49\\
110845.28+594137.9 & 0.7476 & 18.457$\pm$0.026 &   8.95$\pm$ 0.14 & 2002-07-   &   5.20$\pm$ 0.40 &1993-11-23 & 1.83 & 12.38 &  1.34\\
110859.29+031127.9 & 3.4587 & 20.136$\pm$0.036 &  10.36$\pm$ 0.15 & 1998-09-   &   7.40$\pm$ 0.50 &1995-02-27 & 1.99 & 14.68 &  7.84\\
111030.44+034833.3 & 1.8653 & 19.435$\pm$0.021 &  15.37$\pm$ 0.13 & 1998-07-   &  11.20$\pm$ 0.50 &1995-02-27 & 2.28 & 14.23 &  5.52\\
111221.82+003028.5 & 0.5234 & 19.417$\pm$0.029 &   8.88$\pm$ 0.14 & 1998-08-   &   6.70$\pm$ 0.50 &1995-02-27 & 2.25 & 12.57 &  1.54\\
114856.79+555827.3 & 0.9616 & 19.267$\pm$0.074 &  21.88$\pm$ 0.14 & 1997-05-   &  15.10$\pm$ 0.90 &1993-11-23 & 2.49 & 13.70 &  3.69\\
120331.10$-$014111.6 & 1.8316 & 18.308$\pm$0.015 &  11.84$\pm$ 0.14 & 1998-08-   &   8.10$\pm$ 0.50 &1995-02-27 & 1.69 & 14.14 &  5.16\\
121143.52+013011.2 & 2.5902 & 18.520$\pm$0.041 &  18.36$\pm$ 0.15 & 1998-07-   &  14.40$\pm$ 0.60 &1995-02-27 & 1.91 & 14.55 &  7.08\\
121440.07+600330.9 & 1.4863 & 18.589$\pm$0.026 &  34.02$\pm$ 0.14 & 2002-07-   &  17.50$\pm$ 0.70 &1993-11-23 & 2.29 & 13.76 &  3.86\\
121446.06+532023.5 & 2.1472 & 19.686$\pm$0.024 &  13.02$\pm$ 0.22 & 1997-05-05 &   8.90$\pm$ 0.90 &1993-11-15 & 2.04 & 14.34 &  6.01\\
121729.29+060750.8 & 2.0945 & 19.587$\pm$0.019 &  24.49$\pm$ 0.17 & 2000-02-   &  13.50$\pm$ 0.60 &1995-02-27 & 2.44 & 14.43 &  6.46\\
121729.84$-$004715.7 & 1.3371 & 20.205$\pm$0.049 &  18.83$\pm$ 0.14 & 1998-08-   &  14.80$\pm$ 0.60 &1995-02-27 & 2.71 & 13.83 &  4.08\\
121916.76+623026.1 & 3.0559 & 19.553$\pm$0.025 &   8.40$\pm$ 0.14 & 2002-07-   &   6.40$\pm$ 0.40 &1993-11-23 & 2.00 & 13.61 &  3.43\\
122400.78+005919.9 & 1.4956 & 19.475$\pm$0.020 &   5.52$\pm$ 0.14 & 1998-08-   &   2.40$\pm$ 0.50 &1995-02-27 & 1.80 & 13.84 &  4.11\\
122705.72+631533.2 & 1.5937 & 19.672$\pm$0.031 &   5.02$\pm$ 0.13 & 2002-08-   &   3.20$\pm$ 0.40 &1993-11-23 & 1.97 & 12.87 &  1.95\\
122757.23+101410.7 & 1.2924 & 18.145$\pm$0.032 &  7.24$\pm$0.14   & 200001   &  2.90$\pm$0.60  &1995-02-07 & 1.40 & 13.53 &  4.33\\
122819.25+023229.3 & 3.1479 & 20.638$\pm$0.032 & 111.11$\pm$ 1.72 & 1998-09-   &  60.00$\pm$ 1.80 &1995-02-27 & 3.19 & 15.82 & 18.80\\
122956.17$-$012910.6 & 0.9991 & 19.673$\pm$0.021 &   6.55$\pm$ 0.15 & 1998-08-   &   4.50$\pm$ 0.40 &1995-02-27 & 2.11 & 13.23 &  2.56\\
123132.37+013814.0 & 3.2286 & 19.205$\pm$0.026 &  11.51$\pm$ 0.81 & 1998-07-   &   6.30$\pm$ 0.40 &1995-02-27 & 1.85 & 14.90 &  9.26\\
123628.79+565156.4 & 2.5105 & 20.260$\pm$0.025 &   8.36$\pm$ 0.14 & 1997-05-   &   6.40$\pm$ 0.40 &1993-11-23 & 2.30 & 14.19 &  5.36\\
123932.75+044305.3 & 1.7621 & 20.481$\pm$0.025 & 426.95$\pm$ 0.14 & 2000-02-   & 353.80$\pm$10.60 &1995-02-27 & 4.14 & 15.07 & 10.56\\
125014.30+621032.4 & 1.9053 & 19.615$\pm$0.020 &  11.54$\pm$ 0.15 & 2002-07-26 &   9.10$\pm$ 0.50 &1993-11-23 & 2.16 & 13.19 &  2.49\\
125414.27+024117.5 & 1.8405 & 18.816$\pm$0.020 &   6.42$\pm$ 0.14 & 1998-07-   &   4.40$\pm$ 0.40 &1995-02-27 & 1.62 & 13.90 &  4.29\\
131728.65+060046.5 & 2.6095 & 19.127$\pm$0.020 &  76.59$\pm$ 0.14 & 2000-02-   &  37.40$\pm$ 1.20 &1995-02-27 & 2.92 & 15.22 & 11.80\\
131906.47+493152.9 & 1.9322 & 18.887$\pm$0.019 &  13.65$\pm$ 0.13 & 1997-04-17 &  10.20$\pm$ 0.50 &1995-03-12 & 2.03 & 14.58 &  7.27\\
135213.31+581536.8 & 3.1136 & 19.024$\pm$0.028 &  13.54$\pm$ 0.13 & 2001-03-   &   9.60$\pm$ 0.50 &1993-11-23 & 2.09 & 14.07 &  4.88\\
135341.72+431052.5 & 1.1136 & 17.274$\pm$0.023 &  23.15$\pm$ 0.14 & 1997-02-20 &  18.50$\pm$ 0.70 &1995-03-12 & 1.70 & 14.19 &  5.37\\
142730.43+545601.6 & 1.7533 & 17.763$\pm$0.024 &  32.85$\pm$ 0.15 & 1997-03-   &  24.10$\pm$ 0.80 &1993-11-23 & 1.96 & 14.50 &  6.81\\
143540.20+024226.4 & 2.1812 & 19.945$\pm$0.023 &  57.82$\pm$ 0.14 & 1998-07-   &  45.30$\pm$ 1.40 &1995-02-27 & 3.15 & 14.87 &  9.04\\
143623.97+031155.5 & 1.7979 & 19.319$\pm$0.022 &  16.63$\pm$ 0.33 & 1998-07-   &  10.50$\pm$ 0.50 &1995-02-27 & 2.14 & 14.35 &  6.09\\
145002.45+001629.4 & 0.9573 & 19.145$\pm$0.015 &  13.82$\pm$ 0.17 & 1998-07-   &   9.40$\pm$ 0.50 &1995-02-27 & 2.24 & 13.54 &  3.25\\
145207.94+025019.4 & 2.4330 & 20.262$\pm$0.031 &  11.17$\pm$ 0.13 & 1998-07-   &   6.30$\pm$ 0.50 &1995-02-27 & 2.26 & 14.57 &  7.21\\
150205.38+604534.3 & 1.2986 & 19.690$\pm$0.021 &  44.89$\pm$ 0.15 & 2002-07-   &  35.20$\pm$ 1.10 &1993-11-23 & 2.81 & 13.38 &  2.89\\
151002.93+570243.3 & 4.3087 & 22.055$\pm$0.065 & 248.07$\pm$ 0.13 & 1997-05-   & 202.00$\pm$ 6.10 &1993-11-23 & 3.72 & 16.12 & 23.64\\
153559.67+583430.9 & 2.1813 & 18.620$\pm$0.025 &   6.70$\pm$ 0.13 & 2002-06-   &   4.10$\pm$ 0.40 &1993-11-23 & 1.65 & 13.37 &  2.87\\
153703.94+533219.9 & 2.4035 & 18.133$\pm$0.024 &   9.28$\pm$ 0.14 & 1997-05-   &   7.10$\pm$ 0.40 &1993-11-15 & 1.50 & 14.18 &  5.34\\
162548.79+264658.7 & 2.5177 & 17.340$\pm$0.017 &  10.12$\pm$ 0.13 & 1995-12-17 &   6.10$\pm$ 0.40 &1995-04-16 & 1.33 & 15.92 & 20.29\\
162816.95+351023.6 & 0.7151 & 18.713$\pm$0.019 &  15.91$\pm$ 0.14 & 1994-07-03 &   9.40$\pm$ 0.50 &1995-04-16 & 2.23 & 14.65 &  7.64\\
163915.80+412833.7 & 0.6900 & 19.133$\pm$0.016 &  89.23$\pm$ 0.16 & 1994-09-02 &  73.80$\pm$ 2.30 &1995-04-16 & 3.10 & 15.19 & 11.60\\
164602.25+432156.3 & 2.9102 & 20.603$\pm$0.025 &   6.21$\pm$ 0.14 & 1997-02-20 &   3.80$\pm$ 0.40 &1995-03-12 & 2.20 & 14.93 &  9.45\\
164952.90+325815.1 & 0.7109 & 18.530$\pm$0.013 &  43.58$\pm$ 0.12 & 1995-10-14 &  33.60$\pm$ 1.10 &1995-04-16 & 2.55 & 15.23 & 11.91\\
171535.96+632336.0 & 2.1818 & 18.584$\pm$0.019 &  52.48$\pm$ 0.14 & 2002-08-   &  35.90$\pm$ 1.10 &1995-04-02 & 2.33 & 14.31 &  5.88\\
210757.67$-$062010.6 & 0.6456 & 17.496$\pm$0.014 &  19.21$\pm$ 0.14 & 1997-02-   &  12.40$\pm$ 0.60 &1993-09-20 & 1.86 & 13.30 &  2.70\\
230845.85+011201.3 & 3.0559 & 20.146$\pm$0.025 &   8.06$\pm$ 0.13 & 1995-10-16 &   5.50$\pm$ 0.40 &1993-11-15 & 2.33 & 15.02 & 10.13\\

\enddata
\end{deluxetable}
\end{document}